\documentclass[twocolumn,tighten,numberedappendix]{aastex631}
\usepackage{amsmath}

\def\simlt{\lower.5ex\hbox{$\; \buildrel < \over \sim \;$}}
\def\simgt{\lower.5ex\hbox{$\; \buildrel > \over \sim \;$}}

\def\beq{\begin{equation}}
\def\eeq{\end{equation}}
\def\ba{\begin{eqnarray}}
\def\ea{\end{eqnarray}}

\def\M{{\cal M}}

\def\n{\tilde{n}}
\def\B{{\cal B}}
\def\D{\tilde{\Delta}}

\def\E{{\cal E}}
\def\N{{\cal N}}
\def\I{{\cal I}}

\def\F{{\cal F}}

\def\tt{t_1}

\def\I{{\cal I}}

\def\P{{\cal P}}

\def\N{{\cal N}}

\newbox\grsign \setbox\grsign=\hbox{$>$} \newdimen\grdimen \grdimen=\ht\grsign
\newbox\simlessbox \newbox\simgreatbox \newbox\simpropbox
\setbox\simgreatbox=\hbox{\raise.5ex\hbox{$>$}\llap
     {\lower.5ex\hbox{$\sim$}}}\ht1=\grdimen\dp1=0pt
\setbox\simlessbox=\hbox{\raise.5ex\hbox{$<$}\llap
     {\lower.5ex\hbox{$\sim$}}}\ht2=\grdimen\dp2=0pt
\setbox\simpropbox=\hbox{\raise.5ex\hbox{$\propto$}\llap
     {\lower.5ex\hbox{$\sim$}}}\ht2=\grdimen\dp2=0pt
\def\simgt{\mathrel{\copy\simgreatbox}}
\def\simlt{\mathrel{\copy\simlessbox}}

\begin{document}

\title{Upper Field Strength Limit of Fast Radio Bursts}

\author{Yu Zhang and Hui-Chun Wu}

\thanks{huichunwu@zju.edu.cn}

\affiliation{Institute for Fusion Theory and Simulation and Department of Physics,
Zhejiang University, Hangzhou 310027, China}

\begin{abstract}
Fast radio bursts (FRBs) are cosmological radio transients with
unclear generation mechanism. Known characteristics such as their luminosity, duration, spectrum and repetition rate, etc. suggest that FRBs are powerful coherent radio signals at GHz frequencies, but the status of FRBs near source remain unknown. As an extreme astronomical event, FRBs should be accompanied by energy-comparable
or even more powerful x/$\gamma$-ray counterparts.
Here, particle-in-cell simulations of ultrastrong GHz radio pulse interaction with GeV photons show that at $\gtrsim3\times10^{12}$V/cm field-strengths, quantum cascade can generate dense pair plasmas, which greatly dampen the radio pulse. Thus, in the presence of GeV photons in the source region, GHz radio pulses stronger than $3\times10^{12}$V/cm cannot escape. This result indicates an upper field-strength limit of FRB at the source.
%%Investigation on the toleration of $\gtrsim10^{12}$V/cm
%%radio waves for potential plasma/beam emitters in neutron-star scenerios
%%should be helpful for distinguishing the precise mechanism of FRBs.
\end{abstract}

 \keywords{
  Radio bursts(1339); Radio transient sources (2008); Plasma astrophysics (1261)
}

% \maketitle

%#####################################################################

 \section{Introduction}

Since first discovered in 2007 \citep{Lorimer1}, fast radio bursts
(FRBs) have been recognized as real astronomical events and gain
 much research interest \citep{Katz1,Platts,Zhang1}. Although event reports
and theoretical models of FRBs have exploded over the past decade, the
origin of FRBs remains unclear. Due to large dispersion measures
with hundreds or even thousands of $\textrm{cm}^{-3}$pc, these radio
transients have cosmological origins \citep{Xu}, which has been confirmed
by several events with located host galaxies \citep{Thornton,Petroff}.
Therefore FRBs can be served as novel probes for interstellar and intergalactic
matters \citep{Prochaska,Macquart}.

Assuming an isotropic emission, the luminosity of FRBs ranges from $10^{38}$
to $10^{45}$erg/s \citep{Thornton,Zhang2}, many orders of magnitude more
powerful than radio pulsars. Meanwhile, the ultrahigh brightness temperature
$\sim10^{35}$K \citep{Katz1,Zhang1} indicates that FRB radiations
must be coherent. The millisecond duration implies that the source
is limited to hundreds of kilometers in size, which points to compact
objects in universe, such as white dwarfs, neutron stars, or black holes.
The observed FRB200428 \citep{Bochenek,CHIME1,Lin} in the Milky
Way associated with a hard X-ray burst from magnetar SGR1935+2154,
suggest magnetars can generate FRBs.

Many models of FRBs \citep{Katz1,Platts,Zhang1} have been proposed.
Possible FRB sources are located in- or out-side of the magnetospheres of neutron stars.
In the magnetosphere, the radiation mechanisms include plasma maser emissions
from relativistic plasmas or plasma instabilities \citep{Lyubarsky1},
and curvature radiation of charged bunches \citep{Kumar,Katz2,Yang1,Lu}.
Outside of the magnetosphere, relativistic shocks driven by outflows from
neutron stars may also induce FRBs \citep{Lyubarsky2,Waxman,Metzger,Beloborodov}.
Although the emission region of the FRBs in neutron stars is still
being debated, most theoretical models show that FRBs are accompanied by energy comparable or even more powerful counterparts in x and $\gamma$ rays bands of keV-to-TeV \citep{Chen}. The observed x-rays from FRB200428 are four orders of magnitude more energetic than the radio emission.

FRBs have been detected in the range of 0.3-8 GHz \citep{Zhang1}, with
a bandwidth of hundreds of MHz limited by the detection band of radio
telescopes. The coherent GHz radiation implies an emitter in the sub-meter
scale. In the immediate vicinity of the emitters, FRB corresponds to an extremely
strong microwave. Research on this extreme microwave is rare \citep{Wu0}.
In this paper we simulate the interaction between ultrastrong GHz radio waves
and GeV gamma photons. It is found that at field-strength
of $3\times10^{12}$V/cm, dense pair plasmas are produced by quantum
cascades that significantly dampen the radio pulses. This process should
occur in the FRB emission region and constrains the radiation intensity
near the emitters.

\section{Interaction of strong radio wave and high-energy particles}
\subsection{Estimation of the FRB field strength near the source}
The FRB energy can be expressed as
 $W=d\Omega R^{2}cT\varepsilon_{0}E^{2}$,
where $d\Omega$ is the solid angle of the FRB emission cone, $E$
is the field amplitude at a distance $R$ from the source, $T\approx1$ms
is the duration, $c$ is the light speed, and $\varepsilon_{0}$ is
the vacuum permittivity. For isotropic emission ($d\Omega=4\pi$),
the energy range is estimated to be $W_{i}=10^{35}\sim10^{42}$erg, so that the
field strength $E$ can be obtained from

\begin{equation}
W_{i}=4\pi R^{2} cT \varepsilon_{0}E^{2}.
\end{equation}
It is stressed that here $R$ only refers to a distance from the FRB emitters,
which may be located in the inner or outer magnetospheres of neutron stars.

The blue region in Figure 1 shows the possible $E$ range of FRB as a function
of $R$. The field strength range is $1.7\times10^{9}\sim5.4\times10^{12}$V/cm
for $R=100$ km. The solid line marks the Schwinger field $E_{s}=1.32\times10^{16}$V/cm.
The dashed line is the critical field strength $3\times10^{12}$V/cm
predicted by our simulations, where high-energy photons can trigger
strong quantum cascades and radiation damping.
For a point source, $E$ in Equation (1) will diverge when $R\to 0$. The actual emission
zone should consist of many emitters and have a total emission surface
of area $S$. The emission energy has $W=ScT\varepsilon_{0}E^{2}$,
similar to Equation (1). For $S=10^{5}\mathrm{km^{2}}$, $E=1.95\times10^{9}\sim6.14\times10^{12}$V/cm.

\begin{figure}[t]
\centering
\includegraphics[width=0.45\textwidth]{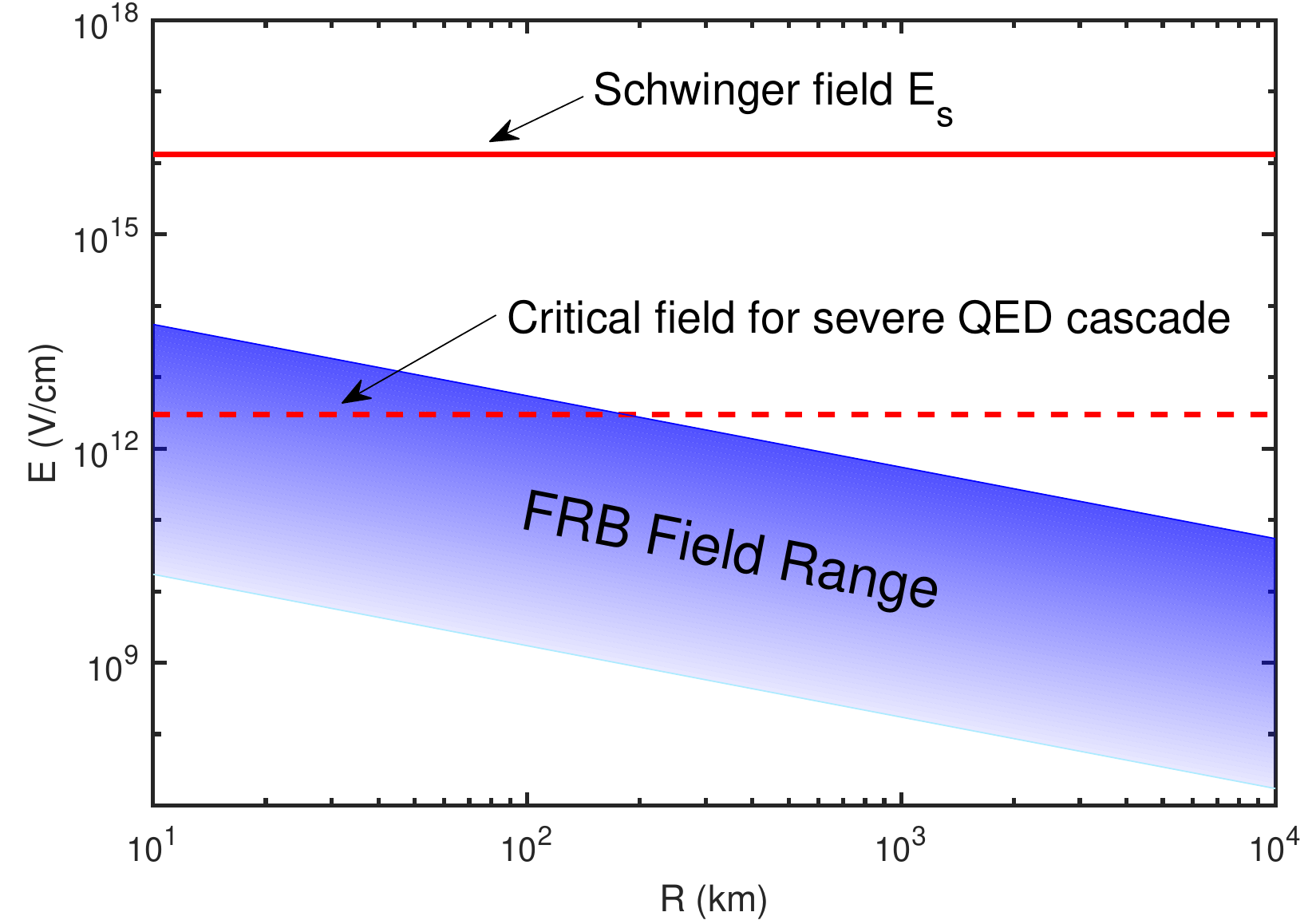}
\label{fig1} \caption{Estimated FRB field strength $E$ as a function of the distance $R$
from sources. }
\end{figure}

\subsection{1D PIC-QED code}

To investigate the interaction between ultrastrong FRBs and gamma photons, we use the one-dimensional particle-in-cell simulation code JPIC1d-QED, where avalanche quantum-electrodynamic (QED) production of positron-electron pairs through the Breit-Wheeler process and nonlinear inverse Compton scattering \citep{Elkina,Erber,Kirk} are included in the existing code JPIC1d \citep{Wu1}. Production of positron-electron pairs and photons are determined by a Monte-Carlo algorithm with
quantum generation rates \citep{Elkina,Wang,Nerush}. A particle-merging
scheme is used to deal with the rapidly increased particles in the
avalanche. To suppress numerical noises typically encountered in PIC-QED
simulations, we adopt a five-point particle interpolation for the positrons/electrons.
The code has been benchmarked for single-electron quantum cascades
in a static magnetic field \citep{Elkina,Anguelov} and reproduced the
results of ultraintense laser QED breakdowns triggered by a single
electron \citep{Nerush}.

\begin{figure}[t]
\centering
\includegraphics[width=0.45\textwidth]{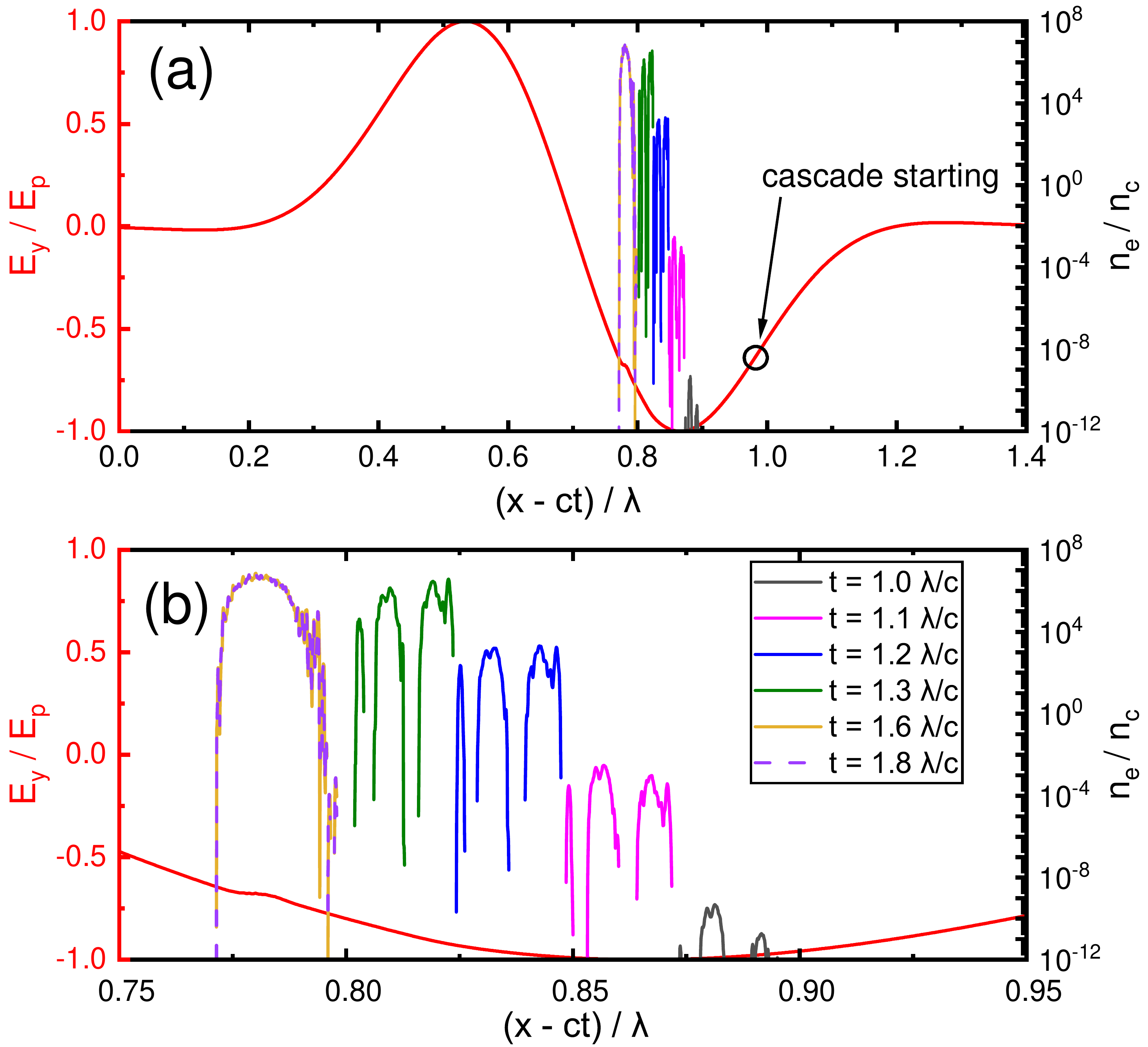}
\caption{(a) Distribution of electron density $n_{e}$ (the positron density is effectively the same) in the comoving frame with the electric field $E_{y}$. (b) An enlargement of the electron densities $n_{e}$ in (a).}
\end{figure}

The Breit-Wheeler process generates electron-positron pairs by
annihilation of gamma photons in intense electromagnetic fields. Inverse
Compton scattering in turn generates gamma photons by the FRB field accelerated relativistic electrons and positrons. Both these effects are measured by (SI units):

\begin{equation}
\chi\simeq\frac{\gamma}{E_{s}}\sqrt{(\boldsymbol{E}+\boldsymbol{v}\times\boldsymbol{B})^{2}-(\boldsymbol{v}\cdot\boldsymbol{E})^{2}/c^{2}}
\end{equation}
where $\gamma$ is the Lorentz factors of electrons/positrons and photons;
$\boldsymbol{v}$ is the particle speed; and $\boldsymbol{E}$ and $\boldsymbol{B}$
are the electric and magnetic fields, respectively. For photons, $\gamma=\epsilon/mc^{2}$ and $\left|\boldsymbol{v}\right|=c$,
$\epsilon$ is the photon energy, and $m$ is the electron rest mass.
Obvious QED cascades can occur for $\chi\simeq0.1$, and massive production of pairs and photons occurs when $\chi\to 1$.

\subsection{Nanosecond-duration radio pulse and GeV gamma photon}

A hundreds-MHz bandwidth implies a coherent time of a few nanoseconds,
thus the millisecond FRBs are expected to contain many coherent nanosecond
subpulses. We now consider the interaction between such nanosecond
subpulse and gamma photons and attempt to find the critical field
strength for substantial quantum cascades within the subcycle of radio
waves. Dense-pair plasmas generated in the subcycle can dampen a great
portion of the entire multiple-cycle pulse \citep{Nerush}. Since FRBs
are highly polarized, we assume the bipolar waveform $E_{y}=E_{0}\exp(-t^{2}/\tau^{2})\sin(\omega_{0}t)$ in our simulations, where $\omega_{0}/2\pi=1$GHz is the central frequency,
and $\tau=0.3\lambda/c$ for wavelength $\lambda\sim 30$cm.
The radio pulse propagates along the $x$ axis and the peak field
strength is $E_{p}=0.636E_{0}$ due to the carrier-envelope phase effect.

Although TeV radiations are also predicted to associate with FRB events, here we
focus on the GeV-level triggering particles, which are not a rigorous
requirement for extreme environments in neutron stars \citep{Becker}.
Quantum cascade is sensitive to the interaction angle between fields
and particles. From Equation (2), one obtains

\begin{equation}
\chi\simeq\frac{\gamma}{E_{s}}(E_{y}-v_{x}B_{z})=\frac{\gamma E_{y}}{E_{s}}(1-\frac{v}{c}\cos\theta)
\end{equation}
where $\theta$ is the angle between the particle velocity
$\boldmath{v}$ and the $x$ axis. The QED effects
are negligible for $\theta=0$ and most pronounced for head-on
collision with $\theta=180^{\circ}$. For $\gamma=2000$ and $\theta=180^{\circ}$,
the field strength for quantum cascades at $\chi=0.1$ is
$E=3.3\times10^{11}$V/cm.

\begin{figure}[t]
\centering
\includegraphics[width=0.48\textwidth]{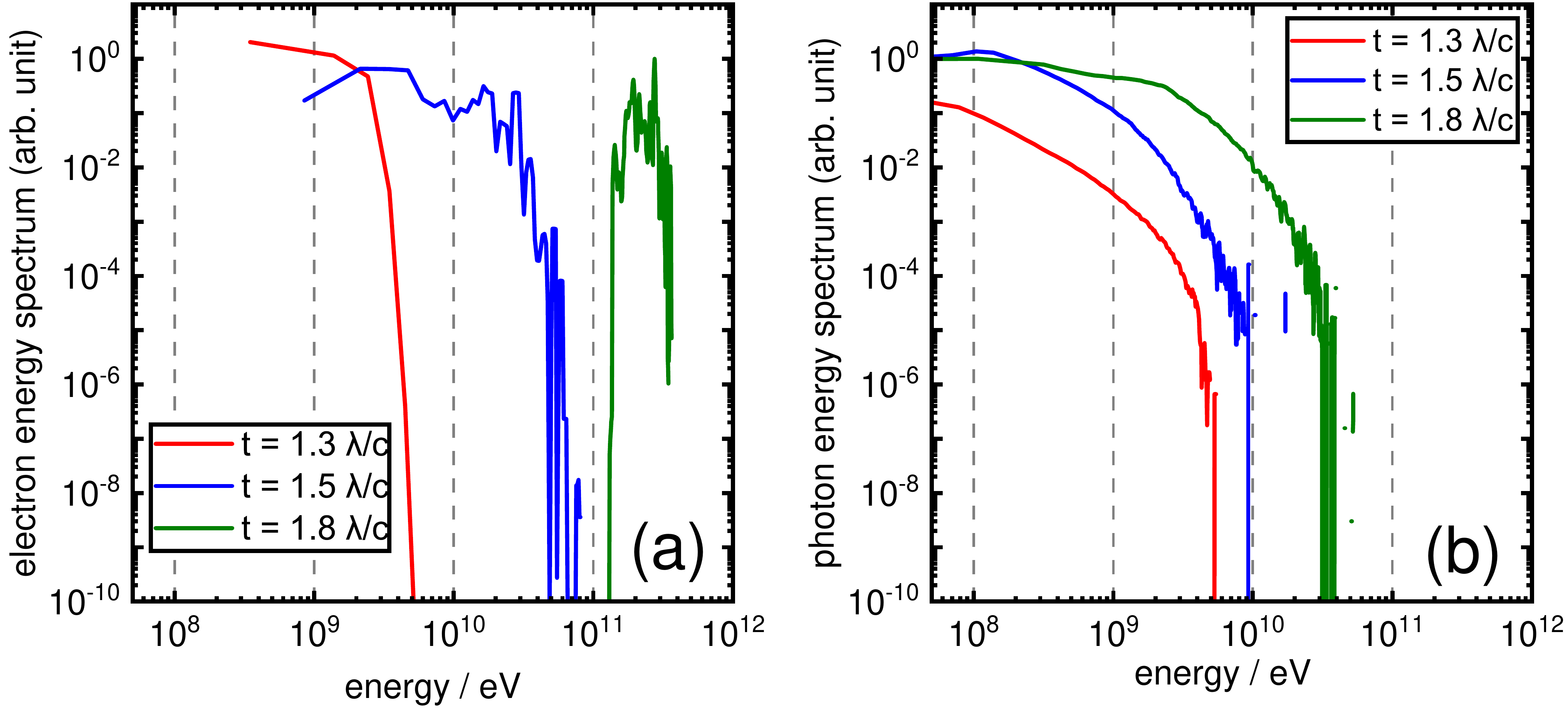}
\caption{Electron energy spectra (a) and photon energy spectra (b) at $t=1.3\lambda/c$,
$1.5\lambda/c$ and $1.8\lambda/c$. Positrons have the almost same
energy spectrum as electrons.}
\end{figure}

It is difficult for the GeV charged particles to trigger quantum cascade in the FRB fields, since the electrons and positrons will be decelerated in the longitudinal direction by the ultrastrong ponderomotive force of FRB fields, then get reflected and copropagate (at $\theta\sim 0$) with the radio waves. Such reflection occurs for the incident particle with energy less than
$\gamma_{rf}\sim a_{0}/4$ \citep{Wu2}, where $a_{0}=eE/mc\omega_{0}$. For 1 GHz radio waves with $E=3.3\times10^{11}$V/cm,
one has $a_{0}=3.1\times10^{6}$ and $\gamma_{rf}\approx7.8\times10^{5}$,
i.e., 0.4 TeV. However, as to be shown below, if the pair particles can emit energetic gamma photons, quantum cascade will be possible since the gamma photons can freely penetrate into the strong FRB fields and annihilate into
electron-positron pairs. .

\begin{figure}[t]
\centering
\includegraphics[width=0.45\textwidth]{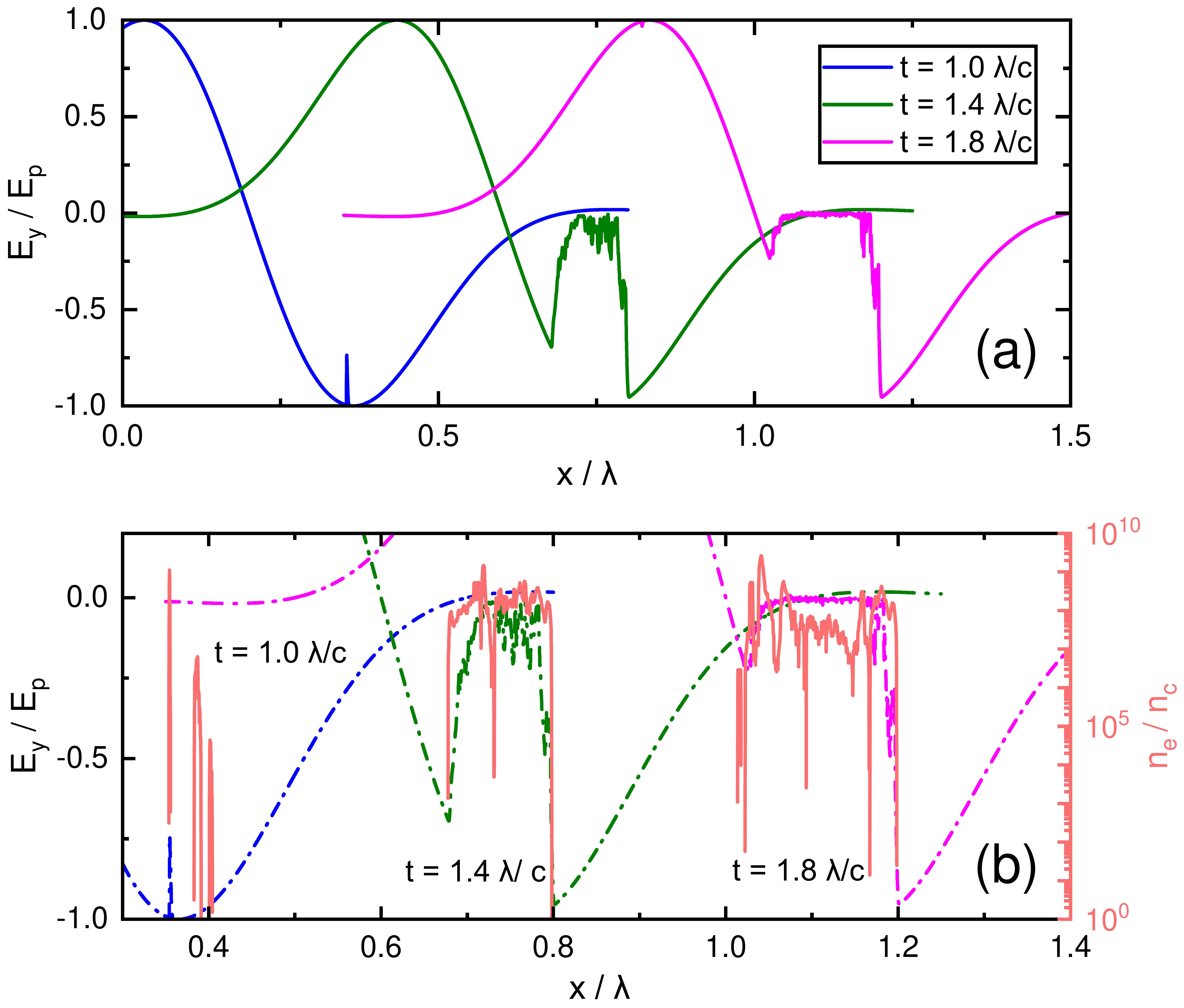}
\caption{(a) Distribution of $E_y$ at
$t=1\lambda/c$, $1.4\lambda/c$, and $1.8\lambda/c$. (b) Enlargement of a region in (a) and $n_e$. The dashed and solid curves represent $E_y$ and $n_e$, respectively.}
\end{figure}

\section{results}

The first simulation is conducted for $E_{p}=2.68\times10^{12}$V/cm, which
corresponds to $a_{0}=eE_{p}/m_{e}c\omega_{0}\simeq2.5\times10^{7}$
and magnetic field of $E_{p}/c\simeq0.89\times10^{10}$ Gauss. The
simulation box is $2\lambda$ in length with a spatio/temporal
resolution of 10000 grids per wavelength/cycle. Ten 1 GeV photons
simultaneously incident to the FRB pulse with $\theta=120^{\circ}$.

Figure 2 shows photon-triggered pair plasma sparks in the comoving frame
of the radio pulse. Obvious QED cascades occur at $E_{y}\approx0.7E_{p0}$,
where annihilation of incident photons generates hundreds of $\chi\sim 0.1$ pair particles, which are in turn violently accelerated by the intense FRB fields to ultra-relativistic energies and emit high-energy photons. The latter are further annihilated into electron-positron pairs by the FRB fields. As shown in Figure 2, three distinct plasma
clumps appear, they grow and finally merge into a plasma sheet within $\sim1$ ns. The plasma density increases exponentially
before the clump merges and saturates at $t=1.6\lambda/c$ with the
peak density of $7.2\times10^{6}n_{c}$. Here, $n_{c}\simeq1\times10^{13}(\mathrm{cm}/\lambda)^{2}\mathrm{cm^{-3}}=1.11\times10^{10}\mathrm{cm^{-3}}$
is the critical density for 1 GHz waves. After $t=1.6\lambda/c$,
the pair plasma sheet comoves with the field, and hence quantum cascades
cease. For an ultra-relativistic field, the plasma density for screening
the field is $\sim a_{0}n_{c}$. The plasma sheet
has a density lower than $a_{0}n_{c}=2.5\times10^{7}n_{c}$ and its thickness is
 $0.01\lambda$. Thus it causes only a small distortion on
the FRB field.

\begin{figure}[t]
\centering
\includegraphics[width=0.45\textwidth]{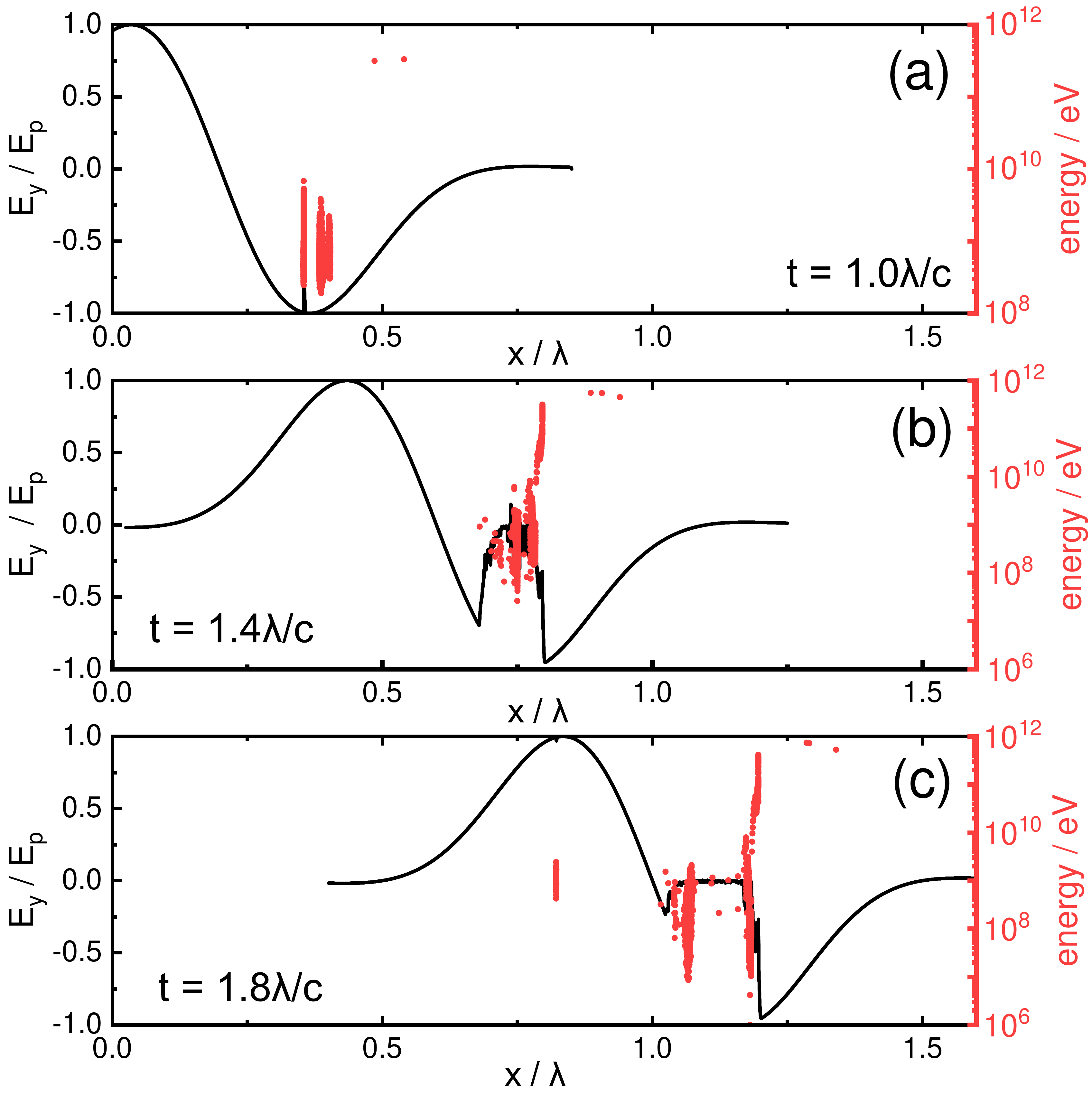}
\caption{Electron energy distribution superposed with the field at
 (a) $t=1\lambda/c$, (b) $t=1.4\lambda/c$ and (c) $t=1.8\lambda/c$.}
\end{figure}

Figure 3 shows the electrons and photons energy spectra. During
the active cascades at $t\sim 1.3\lambda/c$, the particles and photons are synergetic with each
other and have almost the same maximum energy.
However, due to the wave field acceleration, the charged particle energy are always
more concentrated than the photon energy. After the saturation at $t\sim 1.5\lambda/c$, the
charged particles are free from radiation damping, and their energy
quickly overtake that of the photons. They are more monoenergetic than before, and have an average energy $\sim0.2$ TeV at $t=1.8\lambda/c$.

We increase the peak wave field to $E_{p}=3\times10^{12}$ V/cm
with $a_{0}=2.8\times10^{7}$. Figure 4 shows the electric field $E_{y}$
and pair plasma density $n_{e}$. Due to the generated
pair plasmas, partial field screening starts at $t\sim 1\lambda/c$.
Then the plasma density dramatically increases and exceeds the relativistic
critical density $a_{0}n_{c}\sim 2.8\times10^{7}n_{c}$. The field
screening then becomes complete in the plasma with $n_{e}>a_{0}n_{c}$,
and the pair plasma expands towards
the frontside and further grows to the backside. Figure 5 shows the
electron energies distribution. At the front of the
plasma, electrons/positrons are further accelerated along the $x$ axis to higher energies.
According to Equation (3) the QED effects becomes weak and pairs are no longer produced. On the backside, the oscillating charged particles on the vacuum-plasma boundary continues
to collide with the right-going field and produce new pairs. However, Fig. 5(b) and (c) show that due to the continuous radiation damping, the energy of these particles are suppressed, as compared with that of the accelerated particles on the right front. In Figure 5(c), one can see that a plasma spark
appears downstream in the second half cycle. It is triggered by the
emitted photons from the QED-active backside of the pair plasma region.
Such sparks could dampen the FRB field in the more extended space.

\begin{figure}[t]
\centering
\includegraphics[width=0.47\textwidth]{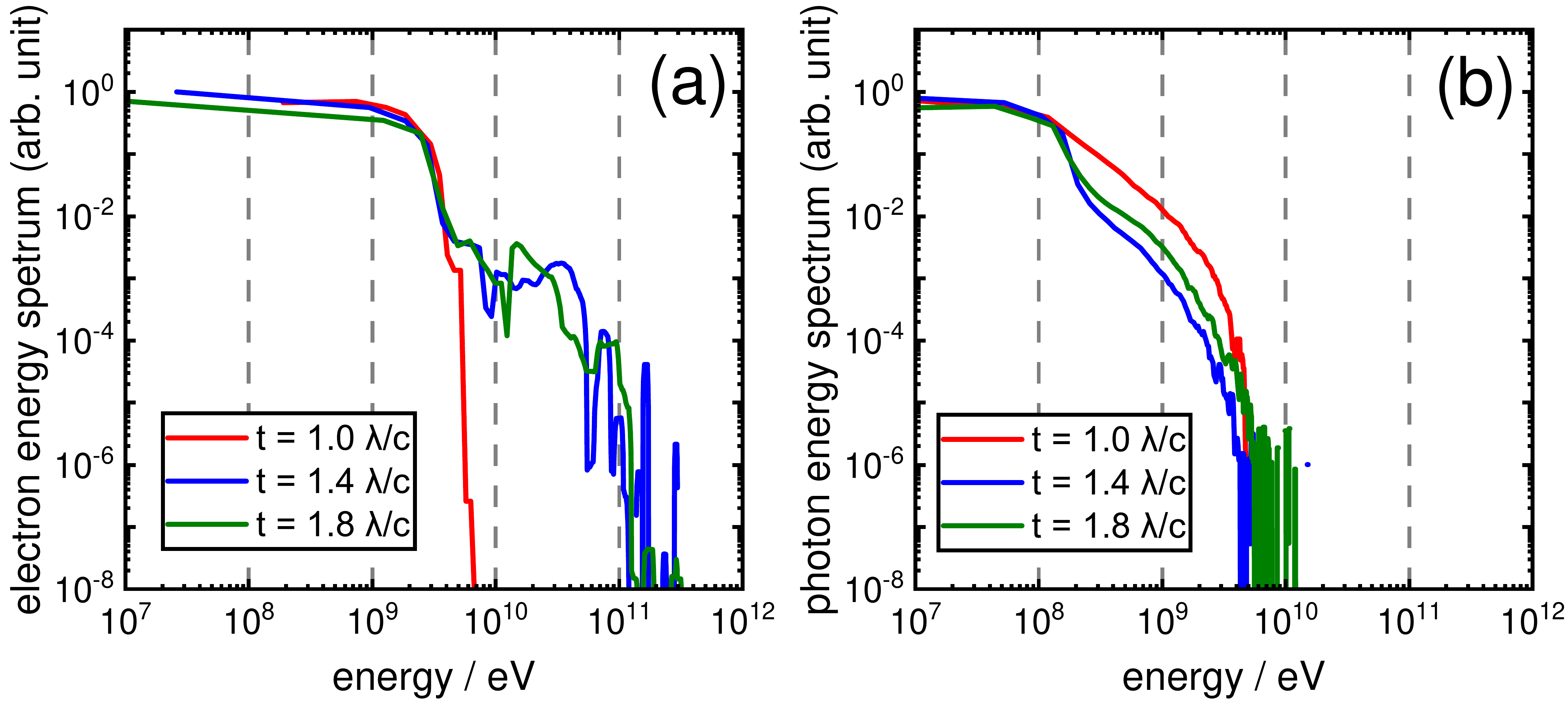}
\caption{Electron energy spectra (a) and photon energy spectra (b)
 at $t=1\lambda/c$, $1.4\lambda/c$ and $1.8\lambda/c$.}
\end{figure}

Energy spectra of electrons and photons are given in Figure 6. During
the active quantum cascades at $t=1\lambda/c$, charged particles
and photons have similar spectra as that in figure 3. After
the complete field screening, the high-energy part of the spectra
at $t=1.4\lambda/c$ and $1.8\lambda/c$ in Figure 6(a) comes from
the field accelerated particles at the plasma front. Depletion of GeV
photons at $1\lambda/c<t<1.4\lambda/c$ can be attributed to photon annihilation and production of pair plasmas.

Gamma photons above 1 GeV are observed to more easily cause the
field depletion. We also carried out the simulations for charged
particles in GeV-TeV range and found that tens GeV are required to trigger
significant radiation damping. In 3D space, pair plasmas will fill the
main radio-field volume, as demonstrated in multi-dimentional
PIC simulation of ultraintense laser breakdown
triggered by a single electron in vacuum \citep{Nerush}. On the other hand, here we have only tens photons interacting with the FRB. Existing models invoke photons with comparable high energy (GeV-TeV) co-existing with the FRBs. The results here indicate that such high-energy photons will fully deplete the FRB.

For FRBs propagating cross the neutron star magnetic field $\boldsymbol{B_{NS}}$, incoherent scattering by magnetospheric plasmas could be significant and gamma rays produced in this process would trigger severe cascades in the FRB\citep{Beloborodov2}. When radio waves propagate along the background magnetic field $B_{NS}$, gamma-photon annihilation due to ${B_{NS}}$ is negligible due to the vanishing term $\boldsymbol{v}\times\boldsymbol{B_{NS}}=0$ in Eq. (2). Therefore, our results are more applicable for FRBs escaping from the polar regions of neutron stars.

As an extremely efficient radio burst, the plasma/beam emitter must be able to coexist with its self-field and radiation field, and will not induce severe field breakdown within both emitter and FRB bodies. According to our results, this requires that the plasma density in or nearby the emitter should be lower than the level of $a_{0}n_{c}$ for less field screening, and the plasma/beam driver could mainly propagate along the radiation direction for negligible pair cascades. The driver can produce high-energy gamma rays along its momentum direction. By Compton backscattering from background plasmas, these gamma rays could turn back to scatter the FRB field as studied here.

\section{conclusion}
We have investigated radiation dampings of FRBs triggered by
high-energy photons in GeV. At the field amplitude
of $3.0\times10^{12}$V/cm, dense pair plasmas are generated within
sub-cycle of these radio transients. The plasma generation and radiation
absorption lead to breakdown of FRBs fields. This energy depletion
can critically limit the field-strength of FRBs around
their emitters. Similar QED effects can also be expected during other stages of FRB propagation.
Our work also implies that QED effects \citep{Philippov,Katz3}
could be indispensable near the FRB emitter.

\medskip

This work was supported by the Strategic
Priority Research Program of Chinese Academy of Sciences (Grant No.XDA17040503).
We thank W. M. Wang and M. Y. Yu for helpful discussions.

\bibliography{ms}

\end{document}